\documentclass[twocolumn,appendixfloats,tighten]{aastex6}
\usepackage{newtxtext,newtxmath}
\usepackage[T1]{fontenc}
\usepackage{ae,aecompl}
\usepackage{amsmath}	
\usepackage{amssymb}	

\usepackage{ctable}
\usepackage{url}
\usepackage{xspace}
\usepackage[normalem]{ulem}
\usepackage{hyperref}
\usepackage[all]{hypcap}
\usepackage{graphicx}
\usepackage{color}

\def\msun{{\rm\,M_\odot}}

\def\msun{{\rm\,M_\odot}}

\newcommand{\be}{\begin{equation}}
\newcommand{\ee}{\end{equation}}

\newcommand{\mbhmax}{M_{BH}^{\rm max}}
\def\h2{${\rm\,H_2}$}

\usepackage{color}

\begin{document}

\title{The Impact of metallicity evolution of the universe on the maximum mass of LIGO binary black holes}

\author{Mohammadtaher Safarzadeh\altaffilmark{1,2} \& Will M. Farr\altaffilmark{3,4}}

\altaffiltext{1}{Center for Astrophysics | Harvard \& Smithsonian, 60 Garden Street, Cambridge MA 02138, USA \href{mailto:msafarzadeh@cfa.harvard.edu}{msafarzadeh@cfa.harvard.edu}}
\altaffiltext{2}{School of Earth and Space Exploration, Arizona State University Tempe AZ 85287, USA}
\altaffiltext{3}{Department of Physics and Astronomy, Stony Brook University, Stony Brook, NY 11794, USA}
\altaffiltext{4}{Center for Computational Astronomy, Flatiron Institute, New York, NY 10010, USA}

\begin{abstract}
We can be biased against observing massive black holes to merge in the local universe as the bounds on the maximum black hole mass ($\mbhmax$)
depends on the assumptions regarding the metallicity evolution of the star forming gas across the cosmic time. 
We investigate the bounds on the metallicity evolution, mass distribution and delay times of the binary black hole sources based on the ten observed events by LIGO.
We parametrize $\mbhmax$ to be a function of metallicity which itself is modeled to evolve with redshift in either a modest or rapid fashion.  
Rapid metallicity evolution models predict a stringent bound of $\mbhmax=44^{+9}_{-5}\msun$, while the bound on $\mbhmax$ in the models with modest metallicity evolution is $\mbhmax=52^{+16}_{-9} \msun$. 
Therefore, inferring $\mbhmax$ from GW data depends on the assumed metal enrichment history of the universe that is not severely constrained at the moment.

\end{abstract}

\section{Introduction}

Detection of binary black holes (BBHs) by LIGO/Virgo has opened a new era in astronomy. 
Much effort has been focused on characterizing the formation scenario of these systems, whether they are born in the field or assembled dynamically. 
There have been studies of the properties of the progenitors of these systems, largely based on the population synthesis models which rely on uncertain physics in large parts.

One of the key questions is whether there exists an upper mass limit for black holes formed through stellar evolution. 
The theoretical models anticipate larger black hole masses to be formed at lower metallicities since the line-driven winds would be quenched, and therefore, a larger mass is available for collapse \citep{Kudritzki:2000cp,Vink:2001cs,Brott:2011fa,Fryer:2012jk}. 
On the other hand, it is believed that pair-instability supernovae (PISN) creates a gap in BH mass distribution, with its location set by the pulsational pair instability supernovae (PPISN) that consequently determine an upper limit on the most massive BHs that can potentially form at the lowest metallicities \citep{Heger:2003ej,Belczynski:2016hj,Yoshida2016,Woosley:2017dj,2018arXiv181013412M,2019arXiv190111136L} due to the mass loss from pulsations pre-supernovae. This leads to the so-called second mass gap between $\approx$ 50 and 135 $\msun$ for BHs formed from stellar core collapse. Given that the space-time volume that LIGO is sensitive to probe scales with the primary mass of the BBH as $m_1^{5/2}$, if there is a cut off at around 50 $\msun$, the evidence for this should be there in the LIGO data.

There has been claims in the literature that the LIGO data so far suggest the presence of a strong upper mass cut for the black holes \citep{Fishbach:2017ic,Talbot:2018cj,Roulet:2019js}.
\citet{Fishbach:2017ic} conclude $\mbhmax=40\msun$, and a power law index of $\alpha<3$ based on about 6 early BBH systems detected by LIGO. 
\citet{Roulet:2019js} arrive at $\mbhmax=41^{+25}_{-10} \msun$ and $\alpha\approx2$ by analyzing the ten observed systems.
LIGO collaboration analysis of the ten events suggests that no more than 1\% of black holes are more massive than 45$\msun$ \citep{Abbottetal:2018vb}.
Moreover, they constrain the power law index of the primary black hole to be $\alpha=1.6^{+1.5}_{-1.7}$(90\% credibility).

One caveat that has been missing in the literature with regards to the $\mbhmax$ is the influence of the metallicity evolution of the universe. 
If black holes close to the $\mbhmax$ limit are born at the lowest metallicities of $\log(Z/Z_{\odot})<-3$, then in order to detect the limit, we need the universe to have 
gone through such low metallicities for enough extended times to provide us with observables. In other words, if PISN is active at $\log(Z/Z_{\odot})<-3$, then if the universe lasted half of its age at such low metallicities, then 
we would have ample evidence for the presence of the upper mass limit. However, if the universe spent only an insignificant lifetime at such low metallicities, then there would have been not much star formation at such 
low metallicities, and therefore, our power to detect the evidence for the presence of such a mechanism would diminish. 

In this paper, we parametrize the distribution of the BBHs with six different parameters, and investigate the constraining power in the ten observed events on them. 
In our model, we tie the maximum black hole mass to the metallicity of the star forming gas, and we parametrize the star forming gas metallicity to evolve either rapidly or slowly with redshift. 
The $\mbhmax$ is considered to be the maximum mass born at zero metallicity and therefore how much time the universe is assumed to have spent at such low metallicities will determine the 
expected birth rate of such massive black holes.

The structure of this paper is as follows: In \S\ref{sec:method} we describe our model in terms of how star forming gas metallicity evolution enters our calculation to set the maximum black hole mass,
and how the inference procedure is carried out. In \S\ref{sec:results} we provide our results, and in \S\ref{sec:summary} we discuss the caveats present in our model. Throughout this paper, we assume Planck 2013 cosmology.

\section{method}\label{sec:method}
\subsection{calculating the merger rate of the BBHs}

The BBH formation rate as a function redshift per comoving volume per source frame time is defined as:
\begin{align}
\frac{dN_{\rm form}}{dm_1 dm_2 dt_f dV_c}=\lambda_{BBH}m_1^{-\alpha} m_2^{-\beta} / C(\alpha,\beta) \psi(z),
\end{align}

where $C(\alpha,\beta)$ is the normalization constant given by:
\be
C(\alpha,\beta)=\int m_1^{-\alpha} m_2^{-\beta} dm_1 dm_2 
\ee
$\psi(z)$ is the cosmic star formation rate density adopted from \citet{Madau:2014gtb}:
\be
\psi(z)=0.015 \frac{(1+z)^{2.7}}{1+[(1+z)/2.9]^{5.6}}\,\, \msun\, {\rm yr^{-1}\, Mpc^{-3}}.
\ee

Here, $\lambda_{BBH}$ is the currently unknown BBH mass efficiency assumed not to evolve with redshift.
The corresponding merger rate is given by:
\be
\frac{dN_{\rm merge}}{dm_1 dm_2 dt_m dV_c}= \int^{\infty}_{t_m(z_m)} P(t_m | t_f) \frac{dN_{\rm form}}{dm_1 dm_2 dt_f dV_c} dt_f
\ee

$P(t_m | t_f)$ is the delay time distribution of the BBHs that sets the probability of merging after $t_m$ of time is past since the formation of the binary. 
We set a minimum delay time, $t_{\rm min}=1$ Myr and impose a maximum delay time of 10 Gyr. 
\be
P(t_m | t_f)= t_m^{-\kappa} / C(\kappa) 
\ee
where $C(\kappa) $ is the normalization constant given by:
\be
C(\kappa)=\int_{t_{\rm min}}^{t_{\rm max}} t_m^{-\kappa} dt.
\ee

Subsequently the merger rate in the detector frame is:
\be
\frac{dN_{\rm merge}}{dm_1 dm_2 dt_d dz} = \frac{dN_{\rm merge}}{dm_1 dm_2 dt_m dV} \frac{dV_c}{dz} \frac{1}{1+z}
\ee
where the redshift derivative of the comoving volume is $dV_c/dz=(4\pi c/H_0)[D_L^2/(1+z)^2E(z)]$, where $D_L$ is the luminosity distance to the source, and $H_0$ is the Hubble constant.

\begin{figure}
\hspace{-0.15in}
\centering
\includegraphics[width=1.0\columnwidth]{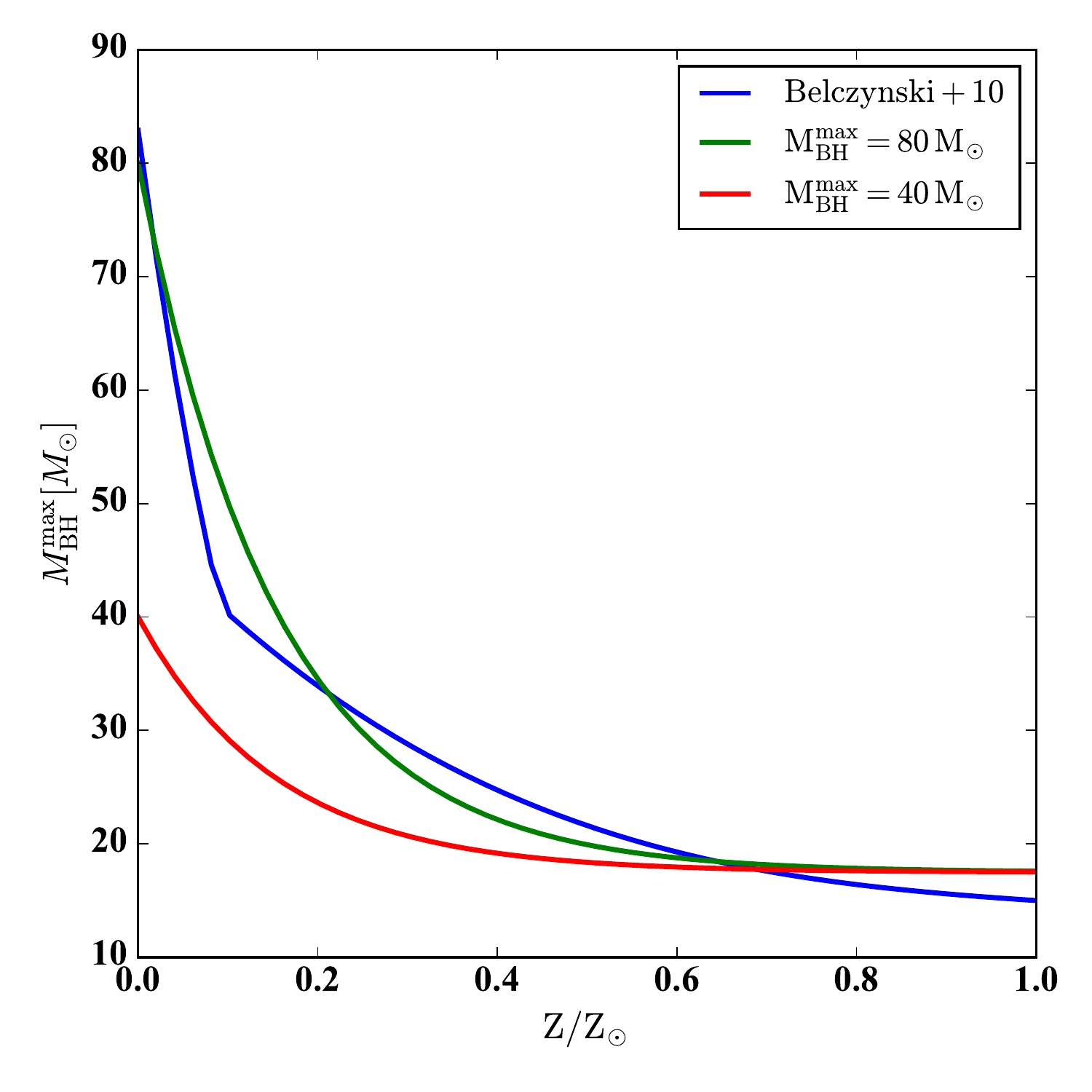}
\includegraphics[width=1.0\columnwidth]{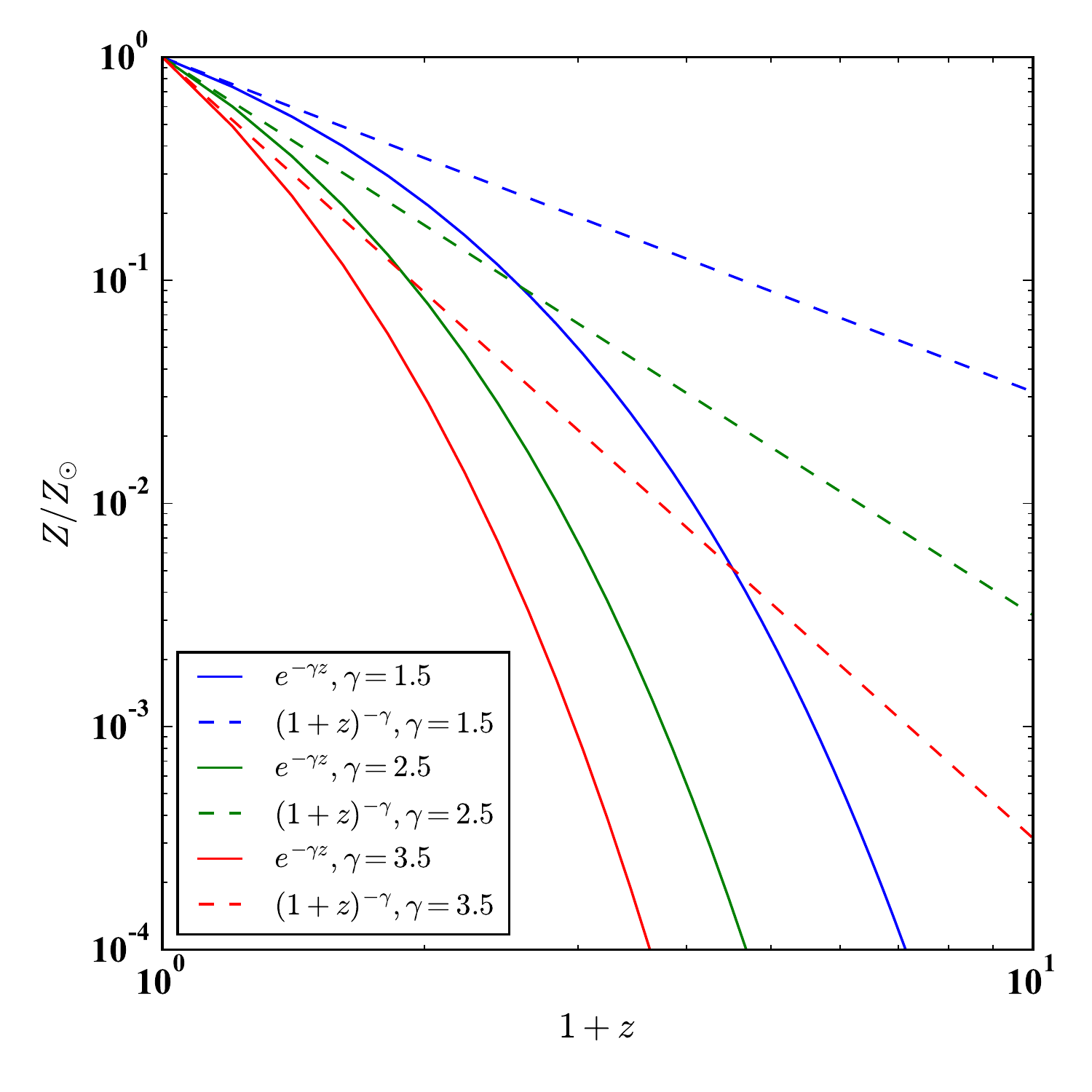}
\caption{\emph{Top panel:} the parametrized maximum black hole mass as a function of metallicity. The blue line shows the results of \citet{Belczynski:2010iw}. The green and red lines show the parametrized 
$\mbhmax$ curves when the $\mbhmax$ is set to 80, and 40 $\msun$ respectively. \emph{Bottom panel:} two different metallicity evolution models adopted in this work, one that drops exponentially with redshift (solid lines), and a power-law model (dashed lines).}
\label{f:fig1}
\end{figure}

In this framework, $M_{\rm min}<m_2<m_1<m_{1}^{\rm max}$, where $M_{\rm min}=5\msun$ and $M_{BH}^{\rm max}$ is set by the metallicity as:
\be
m_{1}^{\rm max}= (M_{BH}^{\rm max}-c) e^{-b Z(z,\gamma)} +c 
\ee
with constants $b=6.5$, and $c=17.5$. This is shown in the top panel of Figure \ref{f:fig1}. This parametrization matches the maximum mass of a blackhole as a function of metallicity as derived in \citet{Belczynski:2010iw} when
we consider the maximum black hole mass to be 80$\msun$. In later series of papers, \citet{Belczynski:2016hj} have included the impact of pair-instability mass loss on black hole binaries which creates 
a second mass gap between 50-150$\msun$ for black holes. Our approach here is whether the presence of PISN could be inferred from the LIGO BBH systems.

$Z(z,\gamma)$ defines the metallicity evolution with redshift that we parametrize in two different ways: The first model is a metallicity evolution in which the metallicity drops exponentially with redshift, i.e., $Z/Z_{\odot}=e^{-\gamma z}$. In the second model the metallicity is modeled as $Z/Z_{\odot}=(1+z)^{-\gamma}$ where the metallicity evolution is much more modest. 
While the impact of metallicity on the LIGO black holes has been explored recently in other works \citep{Kovetz2017,2019arXiv190608136N}, here we explore its impact on the maximum BH mass that LIGO would infer.

The bottom panel of Figure \ref{f:fig1} shows the different models for the metallicity evolution of the universe that is adopted in this work. The metallicity in this work
refers to the star formation rate weighted metallicity of the gas in the galaxies in which the BBHs are born.

From observational perspective, metallicity studies of damped Lyman alpha (DLA) systems at high redshifts suggest a modest evolution at redshifts between $1.5-5$ \citep{Pettini:1997jb,Prochaska:2000dk,Cen:2002fl,Kulkarni:2002eu,Prochaska:2003et,Berg:2016cu}. If the star forming gas metallicity evolves in a similar manner, then low values of $\gamma$ in our parametrization would be the closest model to the observed metallicity evolution.

The {\it observed} BBH merger rate is:
\be
\frac{dN_{\rm obs}}{dm_1 dm_2 dt_d dz}= \frac{dN_{\rm merge}}{dm_1 dm_2 dt_d dz} P_{\text{det}}(m_1,m_2,z),
\ee
where $P_{\text{det}}(m_1,m_2,z)$ is the detection probability of a BBH with masses of $m_1,m_2$, at redshift $z$.
We note that in this work we have assumed the mergers come from the same formation channel, and as such would follow the same $\lambda_{BBH}$ parameter.

\subsection{inference analysis}

To perform our inference analysis, we proceed as follows: Our model has 6 parameters that we fit for $\theta$=($\lambda_{BBH}$,$\alpha,\beta,\gamma,\kappa, M_{BH}^{\rm max}$).
The posterior distribution of these parameters given the data is:
\be
P(\theta | data) = P (data | \theta) P(\theta)
\ee

The prior on our parameter is such that they are bound between $1<\alpha,\beta,\gamma,\kappa<4$, and $ 30<M_{BH}^{\rm max}/\msun<100$.
We approximate $P (data |\theta) $ which we call for brevity $P(d | \theta)$ as follow:
For each BBH event $i$, we have the $P(m_1^i,m_2^i,z^i | d^i )$ from the waveform analysis done by LIGO team.
We have 
\be
P(m_1^i,m_2^i,z^i | d^i )=P(d^i | m_1^i,m_2^i,z^i ) P(m_1^i,m_2^i,z^i )
\ee
and 
\be
P(d^i | \theta) = P(d^i | m_1^i,m_2^i,z^i ) P(m_1^i,m_2^i,z^i | \theta)
\ee
where by combining the last two equations we arrive at 
\be
P(d^i | \theta) \propto P(m_1^i,m_2^i,z^i | d^i ) P(m_1^i,m_2^i,z^i | \theta)
\ee
Therefore, to compute $P(d^i | \theta)$, we draw $N_{\rm sample}$ of $(m_1^j,m_2^j,z^j)^i$ pairs from the posterior $P(m_1^i,m_2^i,z^i | d^i )$ and calculate $P(m_1^j,m_2^j,z^j | \theta)$.
For each event $i$, we have
\be
P(d^i | \theta)=1/N_{\rm sample}\L\sum_{j=0}^{j=N_{\rm sample}} \frac{dN_{\rm merge}}{dm_1 dm_2 dt_d dz}(m_1^j,m_2^j,z^j )^i |_{\theta}
\ee

The posterior distribution from $N_{\rm obs}$ events is:
\be
P(\theta | d) \propto e^{-N_{\rm eff} |_{\theta}} \prod_{i=1}^{i=N_{\rm obs}} P(d^i | \theta) ,
\ee
where $N_{\rm eff}|_{\theta}$ is the expected number of events given $\theta$ defined as :
\be
N_{\rm eff}|_{\theta} = \int_5^{M_{BH}^{\rm max}}  \int_{5}^{m_1}\int_0^{\infty} \int_0^{t_{\rm obs}} \frac{dN_{\rm obs}}{dm_1 dm_2 dt_d dz} |_{\theta}~dm_1 dm_2 dz dt
\ee

Where $t_{\rm obs}$ is the total observing time by LIGO in O1 and O2 runs.

\section{results}\label{sec:results}

\begin{figure*}
\includegraphics[width=1.0\linewidth]{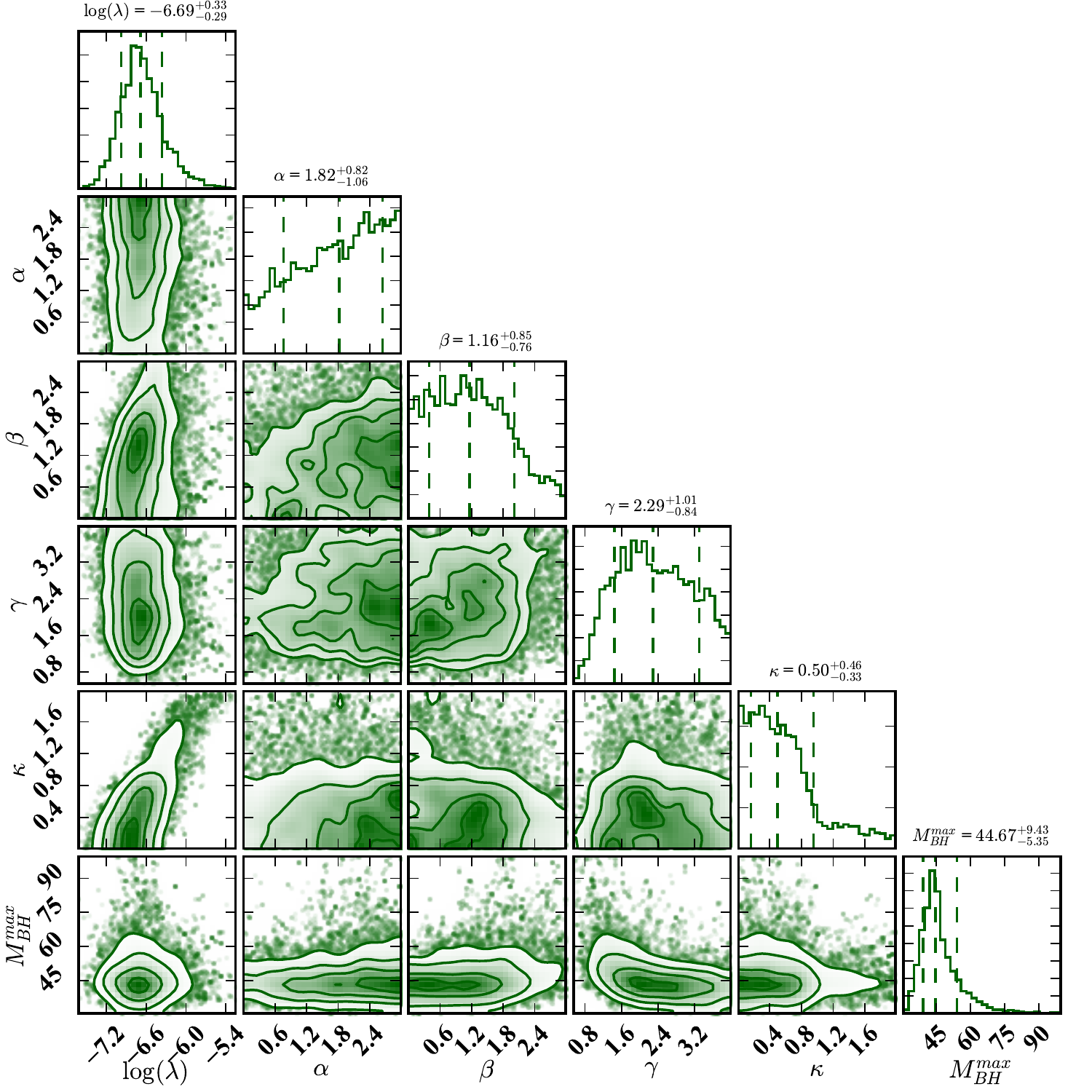}
\caption{Results of MCMC simulation on six parameters in our model by fitting 10 LIGO events in O1 and O2 observing runs. In this model the metallicity evolution is modeled 
as $Z/Z_{\odot}=e^{\gamma z}$.}
\label{f:result_exp}
\end{figure*}

Figure \ref{f:result_exp} shows the posterior distribution for the six parameters of our model when the metallicity evolution is modeled as $Z/Z_{\odot}=e^{-\gamma z}$.
The median BBH efficiency is predicted to be $\approx 2\times10^{-7}/\msun$. This birth rate is robust and is not affected by our metallicity evolution parametrization. 
However, we note that we have assumed $\lambda_{BBH}$ to be constant in this work, but BBH formation is intrinsically tied to this parameter through the wind mass loss. 
So the reader should note that the inferences on $\lambda_{BBH}$ in this work 
is with the prior assumption that $\lambda_{BBH}$ is non-evolving with redshift itself.

While the simulation can not put stringent constraints on $\alpha$, $\beta$, and $\gamma$, one can say large values for $\beta$, and small values of $\gamma$ are disfavored. 
The posterior on $\kappa$ is suggestive of a shallow slope and therefore a preference for long delay times for the BBHs. 
The anti-correlation between the birth efficiency $\lambda_{BBH}$, and $\kappa$ is due to the fact that if a model with long delay times is chosen, then it should be balanced out with lower birth rate efficiency since long delays increase the number density of the BBH mergers in the local universe. Of all the parameters in our model, it is the $\mbhmax$ that is very well constrained to be $\mbhmax=44^{+9}_{-5}$ in this model.

\begin{figure*}
\includegraphics[width=1.0\linewidth]{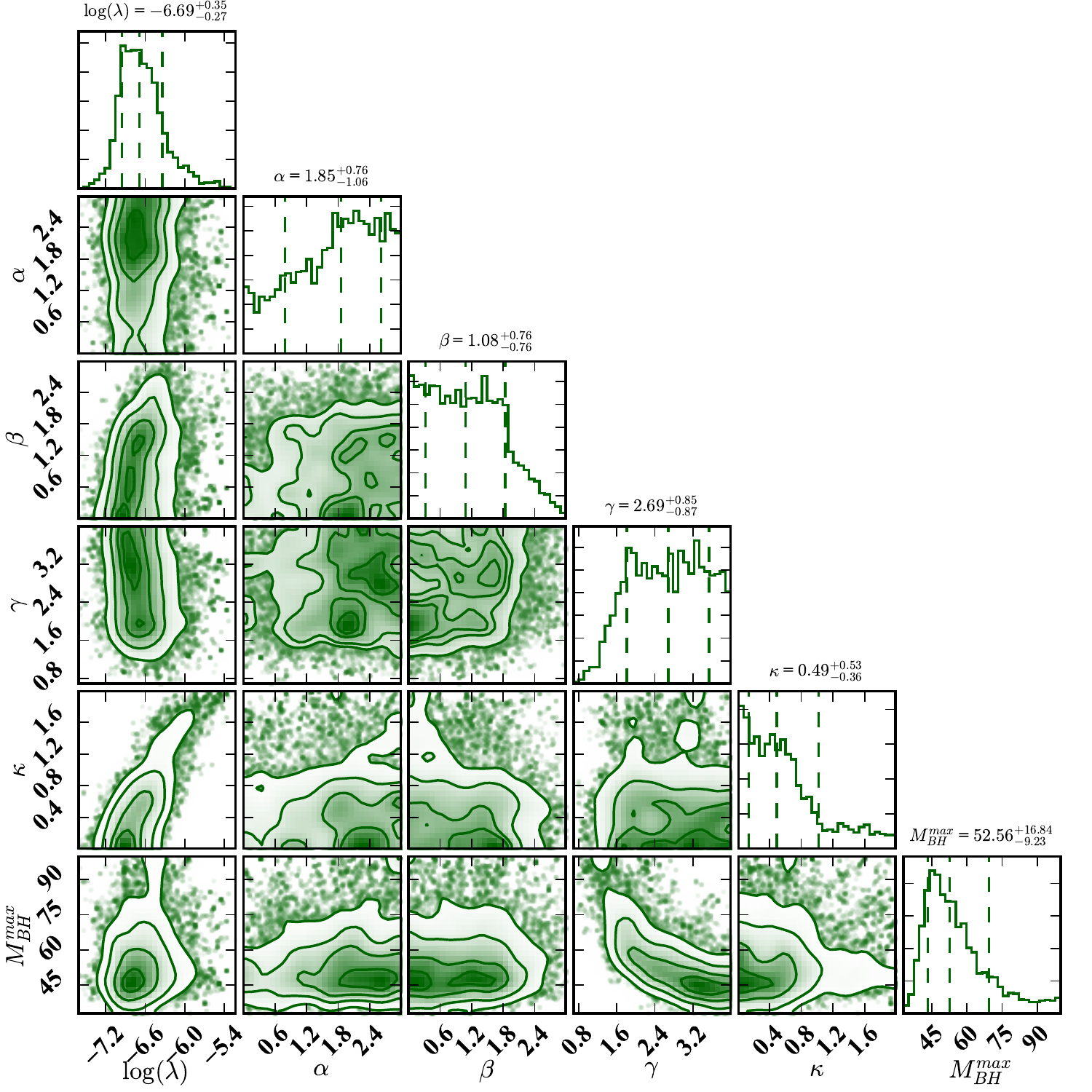}
\caption{Same as in Figure \ref{f:result_exp}  but the metallicity evolution is modeled as $Z/Z_{\odot}=(1+z)^{-\gamma}$. Since in this parametrization the metallicity evolves more gradually with redshift, the impact is evident in the detail of the $\mbhmax$ posterior as larger black hole masses would be allowed in this model. }
\label{f:result_powerlaw}
\end{figure*}

Figure \ref{f:result_powerlaw} shows the same results but for the model with metallicity evolution modeled as $Z/Z_{\odot}=(1+z)^{\gamma}$. It appears that all the parameters expect $\mbhmax$ have the same posterior distribution.
The bounds on the $\mbhmax$ that is less constrained and is $\mbhmax=52^{+16}_{-9}$. Not only the median value is larger, but the upper bound extends to a much larger value.

The impact of the assumptions about the metallicity evolution on the $\mbhmax$ should be understood as follows: In our model, the maximum black hole mass enters our calculation in a non-trivial manner. 
$\mbhmax$ sets the maximum mass that a black hole can have at zero metallicity. If in one model, the metallicity evolution is modest, and barely touches very low metallicities, then to explain the LIGO black holes
one needs to push the $\mbhmax$ to large values to open the room for the model to fit the massive LIGO systems such as GW170729, and GW170823. This is the case when the metallicity evolution follows $(1+z)^{-\gamma}$.
However, if the universe spends much of its cosmic time at very low metallicities, then one can easily explain GW170729 and GW170823 by the star formation at high redshifts. 
The idea can be best seen in the anti-correlation between $\gamma$ and $\mbhmax$ in Figure \ref{f:result_powerlaw}: Large values of $\gamma$ which translate into a faster drop in metallicity, leads to lower values of $\mbhmax$ and vice versa.

\begin{figure*}
\includegraphics[width=1.0\columnwidth]{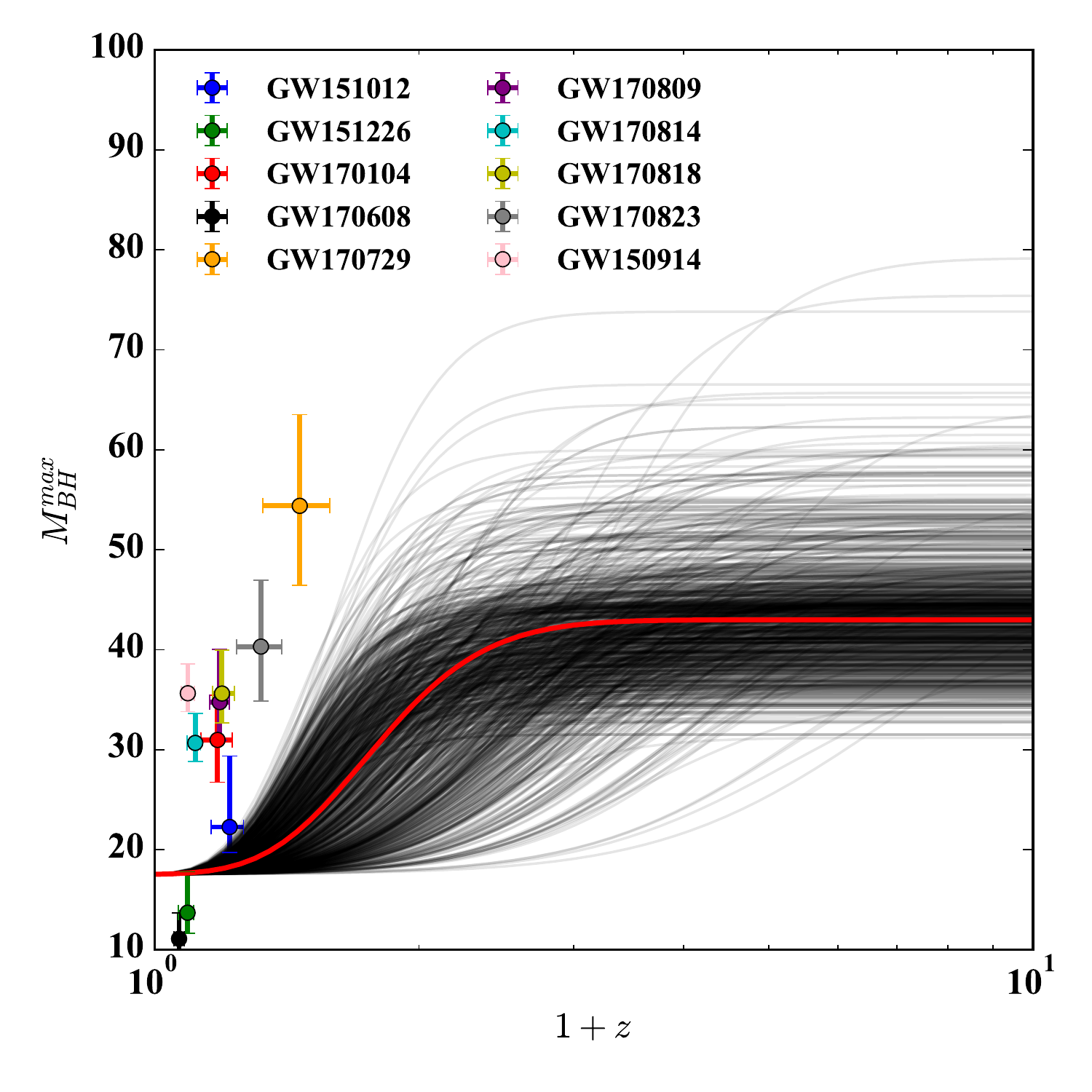}
\includegraphics[width=1.0\columnwidth]{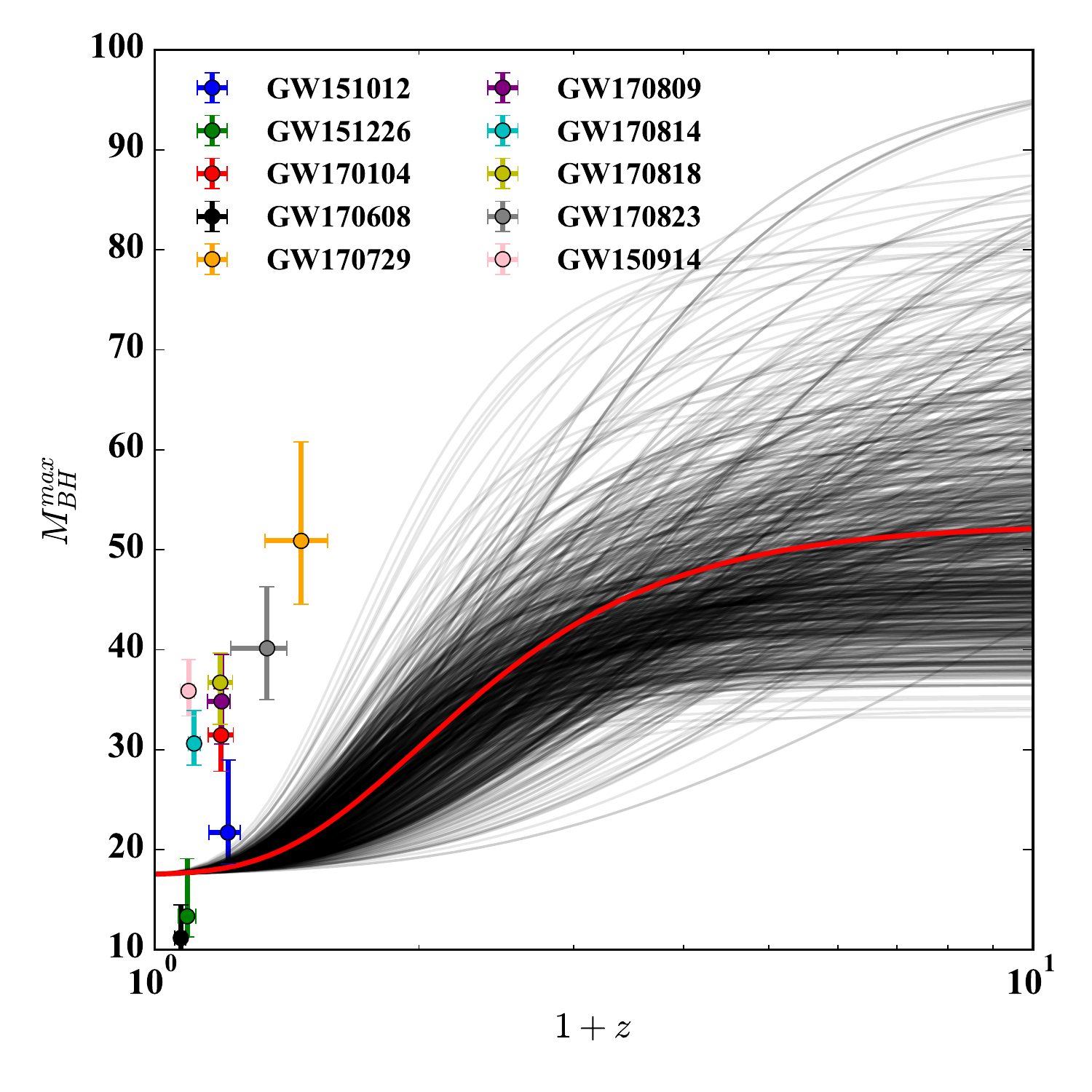}
\caption{{\it Left Panel:} thin black lines show posterior draws from the $\mbhmax$ and $\gamma$ from the model with exponential metallicity evolution with redshift. The thick red line shows the median predicted evolution.
The heavier black hole mass and redshift of the ten observed BBH systems in O1 and O2 observing runs are plotted. The $\mbhmax=44^{+9}_{-5}$ in this model. {\it Right panel:} The same as left panel, but for a metallicity 
evolution parametrized as $\propto (1+z)^{\gamma}$. The bounds on the maximum mass in this model is less constrained and is predicted to be $\mbhmax=52^{+16}_{-9}$.}
\label{f:mbhmax_evol}
\end{figure*}

A different perspective on our results is provided in Figure \ref{f:mbhmax_evol}. In the left panel the thin black lines show posterior draws from the $\mbhmax$ and $\gamma$ from the model with exponential metallicity evolution with redshift. 
The thick red line shows the median predicted evolution. For each of the ten observed BBH systems, we show the bounds on the mass and redshift of the primary (more massive) black hole. 
We are not fitting these data points, we are showing them to be compared to the maximum possible black holes that could be formed above a certain redshift range, such that after a delay time they merger in the local universe. 
The right panel shows the same but for the model with power-law metallicity evolution with redshift.

\begin{figure*}
\includegraphics[width=1.0\columnwidth]{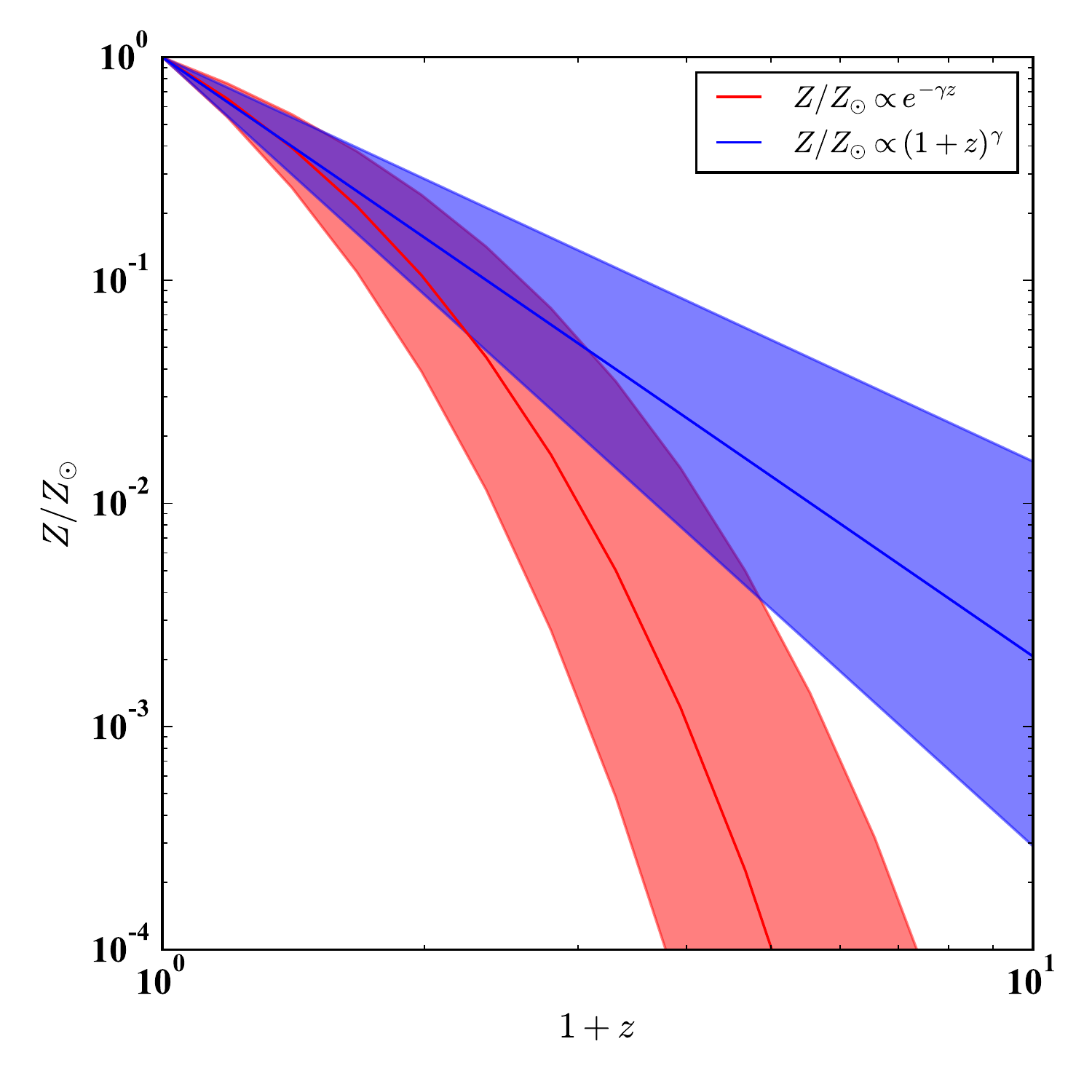}
\includegraphics[width=1.0\columnwidth]{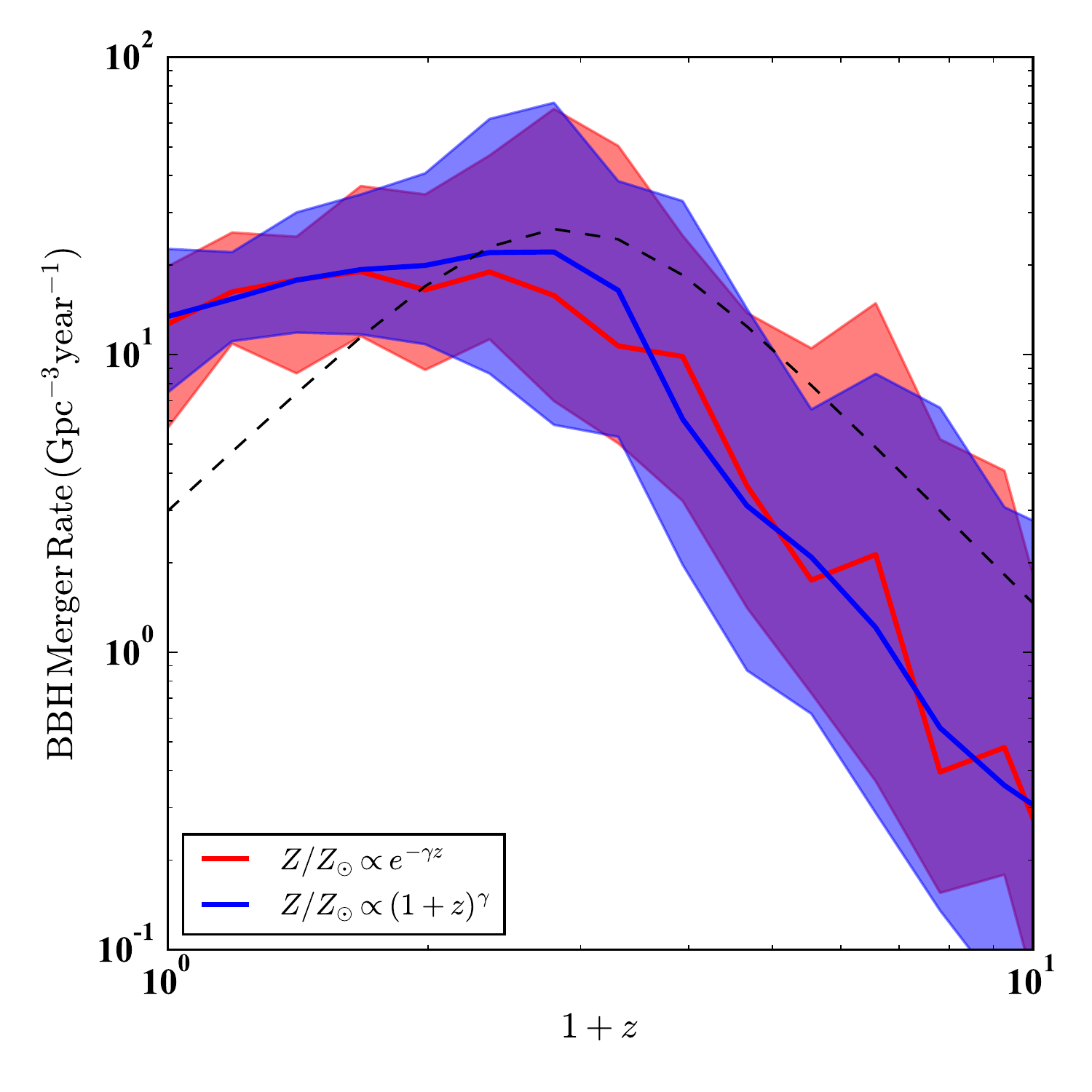}
\caption{{\it Left Panel:} thin black lines show posterior draws of the metallicity evolution in the model with exponential metallicity evolution with redshift. The red line and the shaded region show 
the median and the 16th-84th percentile range. {\it Right panel:} posterior BBH merger rate as a function of redshift for the model with $Z/Z_{\odot}\propto e^{-\gamma z}$ (red shaded region showing the 16th-84th percentile range. The blue line and shaded region show the same for the model with metallicity 
evolution parametrized as $\propto (1+z)^{-\gamma}$. The dashed black line is the $\lambda_{BBH} \psi(z)$, which shows what the merger rate would be if there is no delay time for the BBHs.}
\label{f:metal_evol}
\end{figure*}

The left panel of Figure \ref{f:metal_evol} shows the bounds on the metallicity evolution itself in the two models. 
The solid lines and the shaded region of the same color show 
the median and the 16th-84th percentile range for the each of the metallicity models. The evolution shown with blue is more consistent with the observations of the DLA systems at high redshifts which suggest a modest evolution of their metallicity with redshift \citep{Pettini:1997jb,Prochaska:2000dk,Cen:2002fl,Kulkarni:2002eu,Prochaska:2003et,Berg:2016cu}.

Right panel of Figure \ref{f:metal_evol} shows the
posterior BBH merger rate as a function of redshift for the model with $Z/Z_{\odot}\propto e^{-\gamma z}$ (red shaded region showing the 16th-84th percentile range. The blue line and shaded region show the same for the model with metallicity 
evolution parametrized as $\propto (1+z)^{-\gamma}$. The dashed black line is the $\lambda_{BBH} \psi(z)$, which shows what the merger rate would be if there is no delay time for the BBHs. The different metallicity evolution models did not have a discernible impact on the merger rate of the binaries, and therefore on the maximum mass would be the best probe of the metallicity evolution in this picture.

\section{summary \& discussion}\label{sec:summary}

Our results can be summarized as follows: If maximum black hole mass is set at close to zero metallicity, then in order to infer it from data, it is crucial to have a large part of the cosmic time to have 
a metallicity close to zero to generate BBH systems that can probe the mass limit. In other words, if for example, we lived in a universe in which the metallicity never dropped below 0.1 $Z_{\odot}$, then there would  have been little hope
to constrain a parameter that requires probing metallicities close to $10^{-4}Z_{\odot}$. In our two models, one prescription of the metallicity evolves rapidly with redshift and the other evolves rather smoothly. 
The bounds on the $\mbhmax$ are much more stringent in the model with a rapid drop of metallicity with redshift (i.e., $Z\propto e^{\gamma z}$), compared to the model in which metallicity is modeled as $Z\propto (1+z)^{-\gamma}$ .

Similarly, if we lived in a universe in which the very heavy black holes tend to be born in close binaries, and therefore merger rapidly, then we would be biased against finding them in the local universe. Such a parametrization is not considered in this work, but it would have resulted in the same conclusions that we have reached so far.

Therefore, any claim as to the presence of an upper limit on the $\mbhmax$ should be taken with the caveat that we can be easily biased against them, and the bound on the $\mbhmax$ depends on our 
assumptions with regard to (i) how these systems are born (metallicity range) and how does the universe on average evolve in metallicity, and (ii) whether the more massive systems tend to cluster in a parameter 
space in delay times that we would be biased against them.  

\acknowledgements

This work was supported by the National Science Foundation under grant AST14-07835 and by NASA under theory grant NNX15AK82G as well as a JTF grant. 
MTS is grateful to the Center for Computational Astrophysics for hospitality during the course of this work.


\begin{thebibliography}{}
\expandafter\ifx\csname natexlab\endcsname\relax\def\natexlab#1{#1}\fi

\bibitem[{Abbott et~al . {et~al.}(2018)Abbott et~al ., Collaboration, Abbott,
  Abbott, Abbott, Abraham, Acernese, Ackley, Adams, Adhikari, Adya, Affeldt,
  Agathos, Agatsuma, Aggarwal, Aguiar, Aiello, Ain, Ajith, Allen, Allocca,
  Aloy, Altin, Amato, Ananyeva, Anderson, Anderson, Angelova, Antier, Appert,
  Arai, Araya, Areeda, Ar{\`e}ne, Arnaud, Arun, Ascenzi, Ashton, Aston, Astone,
  Aubin, Aufmuth, AultONeal, Austin, Avendano, Avila-Alvarez, Babak, Bacon,
  Badaracco, Bader, Bae, Baker, Baldaccini, Ballardin, Ballmer, Banagiri,
  Barayoga, Barclay, Barish, Barker, Barkett, Barnum, Barone, Barr, Barsotti,
  Barsuglia, Barta, Bartlett, Bartos, Bassiri, Basti, Bawaj, Bayley, Bazzan,
  B{\'e}csy, Bejger, Belahcene, Bell, Beniwal, Berger, Bergmann, Bernuzzi,
  Bero, Berry, Bersanetti, Bertolini, Betzwieser, Bhandare, Bidler, Bilenko,
  Bilgili, Billingsley, Birch, Birney, Birnholtz, Biscans, Biscoveanu, Bisht,
  Bitossi, \& Bizouard}]{Abbottetal:2018vb}
Abbott et~al ., B.~P., Collaboration, t.~V., Abbott, B.~P., {et~al.} 2018,
  1811.12940

\bibitem[{Belczynski {et~al.}(2010)Belczynski, Bulik, Fryer, Ruiter, Valsecchi,
  Vink, \& Hurley}]{Belczynski:2010iw}
Belczynski, K., Bulik, T., Fryer, C.~L., {et~al.} 2010, The Astrophysical
  Journal, 714, 1217

\bibitem[{Belczynski {et~al.}(2016)Belczynski, Heger, Gladysz, Ruiter, Woosley,
  Wiktorowicz, Chen, Bulik, O'Shaughnesy, Holz, Fryer, \&
  Berti}]{Belczynski:2016hj}
Belczynski, K., Heger, A., Gladysz, W., {et~al.} 2016, Astronomy {\&}
  Astrophysics, A97

\bibitem[{Berg {et~al.}(2016)Berg, Ellison, S{\'a}nchez-Ram{\'\i}rez,
  Prochaska, Lopez, D'Odorico, Becker, Christensen, Cupani, Denney, \&
  Worseck}]{Berg:2016cu}
Berg, T. A.~M., Ellison, S.~L., S{\'a}nchez-Ram{\'\i}rez, R., {et~al.} 2016,
  Monthly Notices of the Royal Astronomical Society, 3021

\bibitem[{Brott {et~al.}(2011)Brott, de~Mink, Cantiello, Langer, de~Koter,
  Evans, Hunter, Trundle, \& Vink}]{Brott:2011fa}
Brott, I., de~Mink, S.~E., Cantiello, M., {et~al.} 2011, Astronomy {\&}
  Astrophysics, A115

\bibitem[{Cen {et~al.}(2002)Cen, Ostriker, Prochaska, \& Wolfe}]{Cen:2002fl}
Cen, R., Ostriker, J.~P., Prochaska, J.~X., \& Wolfe, A.~M. 2002, The
  Astrophysical Journal, 598, 741

\bibitem[{Fishbach \& Holz(2017)}]{Fishbach:2017ic}
Fishbach, M., \& Holz, D.~E. 2017, The Astrophysical Journal Letters, 851, L25

\bibitem[{Fryer {et~al.}(2012)Fryer, Belczynski, Wiktorowicz, Dominik,
  Kalogera, \& Holz}]{Fryer:2012jk}
Fryer, C.~L., Belczynski, K., Wiktorowicz, G., {et~al.} 2012, The Astrophysical
  Journal, 749, 91

\bibitem[{Heger {et~al.}(2003)Heger, Fryer, Woosley, Langer, \&
  Hartmann}]{Heger:2003ej}
Heger, A., Fryer, C.~L., Woosley, S.~E., Langer, N., \& Hartmann, D.~H. 2003,
  The Astrophysical Journal, 591, 288

\bibitem[\protect\citeauthoryear{Kovetz et al.}{2017}]{Kovetz2017} Kovetz E.~D., Cholis I., Breysse P.~C., Kamionkowski M., 2017, PhRvD, 95, 103010

\bibitem[{Kudritzki \& Puls(2000)}]{Kudritzki:2000cp}
Kudritzki, R.-P., \& Puls, J. 2000, Annual Review of Astronomy and
  Astrophysics, 38, 613

\bibitem[{Kulkarni \& Fall(2002)}]{Kulkarni:2002eu}
Kulkarni, V.~P., \& Fall, S.~M. 2002, The Astrophysical Journal, 580, 732

\bibitem[\protect\citeauthoryear{Leung, Nomoto \& Blinnikov}{2019}]{2019arXiv190111136L} Leung S.-C., Nomoto K., Blinnikov S., 2019, arXiv, arXiv:1901.11136

\bibitem[{Madau \& Dickinson(2014)}]{Madau:2014gtb}
Madau, P., \& Dickinson, M. 2014, Annual Review of Astronomy and Astrophysics,
  52, 415

\bibitem[\protect\citeauthoryear{Marchant et al.}{2018}]{2018arXiv181013412M} Marchant P., Renzo M., Farmer R., Pappas K.~M.~W., Taam R.~E., de Mink S., Kalogera V., 2018, arXiv, arXiv:1810.13412
\bibitem[\protect\citeauthoryear{Neijssel et al.}{2019}]{2019arXiv190608136N} Neijssel C.~J., et al., 2019, arXiv, arXiv:1906.08136


\bibitem[{Pettini {et~al.}(1997)Pettini, Smith, King, \&
  Hunstead}]{Pettini:1997jb}
Pettini, M., Smith, L.~J., King, D.~L., \& Hunstead, R.~W. 1997, The
  Astrophysical Journal, 486, 665

\bibitem[{Prochaska {et~al.}(2003)Prochaska, Gawiser, Wolfe, Castro, \&
  Djorgovski}]{Prochaska:2003et}
Prochaska, J.~X., Gawiser, E., Wolfe, A.~M., Castro, S., \& Djorgovski, S.~G.
  2003, The Astrophysical Journal, 595, L9

\bibitem[{Prochaska \& Wolfe(2000)}]{Prochaska:2000dk}
Prochaska, J.~X., \& Wolfe, A.~M. 2000, The Astrophysical Journal, 533, L5

\bibitem[{Roulet \& Zaldarriaga(2019)}]{Roulet:2019js}
Roulet, J., \& Zaldarriaga, M. 2019, Monthly Notices of the Royal Astronomical
  Society, 484, 4216

\bibitem[{Talbot \& Thrane(2018)}]{Talbot:2018cj}
Talbot, C., \& Thrane, E. 2018, The Astrophysical Journal, 856, 173

\bibitem[{Vink {et~al.}(2001)Vink, de~Koter, \& Lamers}]{Vink:2001cs}
Vink, J.~S., de~Koter, A., \& Lamers, H. J. G. L.~M. 2001, Astronomy {\&}
  Astrophysics, 369, 574

\bibitem[{Woosley(2017)}]{Woosley:2017dj}
Woosley, S.~E. 2017, The Astrophysical Journal, 836, 244

\bibitem[\protect\citeauthoryear{Yoshida et al.}{2016}]{Yoshida2016} Yoshida T., Umeda H., Maeda K., Ishii T., 2016, MNRAS, 457, 351


\end{thebibliography}
\end{document}